\documentclass[12pt,preprint]{aastex}
\usepackage{bm}
\newcommand{\dpdp}[2]{\frac{\partial #1}{\partial #2}}
\begin{document}
%%%%%%%%%%%%%%%%%
% (1)TITLE PAGE %
%%%%%%%%%%%%%%%%%
\title{Generation of Seed Magnetic Field around First Stars:\\the
Biermann Battery Effect}
\author{Kentaro Doi\altaffilmark{1} and Hajime Susa\altaffilmark{2}}
\affil{Department of Physics, Konan University, Okamoto, Kobe, Japan}
\altaffiltext{1}{dn121001@center.konan-u.ac.jp}
\altaffiltext{2}{susa@konan-u.ac.jp}
%%%%%%%%%%%%%%%%%%%%%%%%%%%%%%%%%%%
% (2)Abstract  & Subject Headings %
%%%%%%%%%%%%%%%%%%%%%%%%%%%%%%%%%%%
\begin{abstract}
We investigate generation processes of magnetic fields around first stars.
Since the first stars are expected to form anisotropic ionization fronts
 in the surrounding clumpy media, magnetic fields are generated by effects
 of radiation force as well as the Biermann battery effect.
We have calculated the amplitude of magnetic field generated by the effects of
 radiation force around the first stars in the preceding
 paper, in which the Biermann battery effects are not
 taken into account.
In this paper, we calculate the generation of magnetic fields by the Biermann
 battery effect as well as the effects of radiation force, utilizing the radiation hydrodynamics simulations.
As a result, we find that the generated magnetic field strengths are $\sim
 10^{-19}{\rm G}-10^{-17}$G at $\sim 100$pc-1kpc scale mainly by the Biermann
 battery, which is an order of magnitude larger than the results of our previous study. 
We also find that this result is insensitive to various physical
 parameters including the mass of the source star, distance between the
 source and the dense clump, unless we take unlikely values of these parameters.
\end{abstract}
\keywords{early universe---HI\hspace{-.1em}I regions ---radiative transfer --- magnetic fields}

\section{Introduction}
According to the theoretical studies in the last decade, first stars are
expected to be very massive ($\ga 100M_{\odot}$)
\citep[e.g.,][]{Bromm02,NU01,Abel02,Yoshida06a}.
Recent studies which properly address the accretion phase of first star formation
also revealed that the primary star formed in the center of the mini-halo is not very massive but still massive $\ga 10~M_\odot$, 
although significant fraction of first stars are less
massive ($\la 1 M_\odot$)\citep{Stacy10,Clark11a,Clark11b,Greif11}.
In any case, star formation episodes in the very early universe are
different from that in local galaxies.
One of the reasons of this difference is that 
the primordial gas clouds that host the first stars lack heavy
elements, though they are most efficient coolants in interstellar clouds at
$T\la 1000$K.
Because of the lack these coolants,  the temperature of the primordial
gas cloud is kept around $\sim$1000K during its collapse for $n_{\rm
H}<10^{12}{\rm cm^{-3}}$, which is much hotter than the local interstellar
molecular gas clouds. Consequently, the Jeans mass of the collapsing
primordial gas is much larger than that of the interstellar gas, which
leads to the formation of very massive stars\cite[e.g.,][]{omukai00}.

Another important difference between the star formation sites in the
early universe and local molecular clouds is the strengths of magnetic
fields. 
Typical field strength $B\sim 10{\rm \mu G}$ in the
local molecular gas results in the formation of jets from protostars and
regulate the gravitational collapse of cloud cores. 
The effects of magnetic field on the star formation in
the early universe have been studied from theoretical aspects.
First of all, the coupling of the magnetic field with the primordial gas was
studied by a detailed chemical reaction model\citep{maki04,maki07}. They
found that magnetic field is basically frozen-in the primordial gas
during its collapse, differently from the local interstellar
gas\citep[e.g.,][]{nakano_umebayashi86a,nakano_umebayashi86b}. 
Under the assumption of the flux freezing condition, the dynamical importance of
the magnetic field is also investigated by several authors. 
In case the magnetic field is stronger than
$\sim 10^{-9}$G at $n_{\rm H}=10^3{\rm cm}^{-3}$, field strength is
amplified to $\sim 10^{3}$G at $n_{\rm H}=10^{21}{\rm cm}^{-3}$ which is
enough to launch the bipolar outflows, and to suppress the fragmentation
of the accretion disk\citep{machida06,machida08}. \citet{tan_blackman}
estimated the condition for the magnetorotational instability (MRI) to
be activated in the accretion disk around the protostar, by comparing the Ohmic
dissipation time scale with the growth time scale of MRI. They found
that the condition is $B \ga 10^{-2}$G at $n_{\rm H}=10^{15}{\rm
cm}^{-3}$ which corresponds to $B \ga 10^{-10}$G at $n_{\rm H}=10^3{\rm cm}^{-3}$.
We also remark that the turbulent motion powered by the accreting gas
can amplify the initial field strength faster than the simple flux
freezing, although the effects are still under debate.
Thus, the magnetic field could be of importance at the final phase of
the star formation process in primordial gas clouds, if $B\ga
10^{-10}-10^{-9}{\rm G}$ at $n_{\rm H}=10^3{\rm cm}^{-3}$, i.e. at the
initial phase of the collapse of primordial gas in the mini-halos.

In addition, recent theoretical studies suggested that the heating by
the ambipolar diffusion process in star-forming gas clouds
could change the thermal evolution of the prestellar core in case the
field strength is as strong as $10^{-10}$G at IGM comoving densities
\citep{schleicher09,sethi08}.
This process also might leads
to the formation of massive black holes\citep{sethi10} since such
heating can shut down the H$_2$ cooling and open the path of the atomic cooling\citep[e.g.,][]{omukai_yoshii03}.

In any case, it is
important to determine the magnetic field strengths 
in star forming gas clouds in the early universe, in order to
quantify the effects of magnetic fields on the primordial star formation.
In spite of such potential importance, initial seed magnetic field strengths
are still unknown observationally. 
Only the observations on the distortion of the cosmic microwave
background spectrum\citep[e.g.,][]{barrow97,Seshadri09}, and the measurements of the Faraday rotation in the
polarized radio emission from distant quasars
\citep[e.g.,][]{Blasi99,Vallee04} imply rather mild upper limits on
the field strengths in IGM, at the level of $B\sim 10^{-9}$G. 
On the other hand, various theoretical studies predicted that it
is as small as $\la 10^{-18}$G at IGM densities.
For instance, there are models generating the magnetic field by the
Biermann battery\citep{biermann} during 
the structure formation \citep{kulsrud97,xu}. Recent numerical simulation by
\cite{sur10} suggests that strong magnetic field emerges during collapse
of turbulent prestellar cores of primordial gas due to the Biermann
battery and the turbulent dynamo action. 
There is also a number of models 
that the fields are generated just after the big bang\citep[e.g.,][]{turner,ichiki}.

It is also suggested that reionization of the universe inevitably
generates magnetic fields. \citet{gnedin} have shown that the considerable
Biermann battery term arises at the ionization fronts in
their cosmological simulations. 
They predict $\sim 10^{-18}$G at $\delta \rho/\rho \simeq 10^3$. 
It also is suggested that in the neighborhood of luminous sources like
QSOs\citep{langer} or first stars (\citet{ando10}; here after ADS10)
magnetic field could be generated through the momentum transfer process from
ionizing photons to electrons, since $\bm{\nabla}\times{\bm{E}}\ne 0$ is
satisfied at the borders between the shadowed regions and ionized regions.
ADS10 predicted $B\sim 10^{-19}$G at IGM densities at $z=20$, however,
they did not take into account the Biermann battery effect, since they
assume that the gas is isothermal and static.

In this paper, we extend our previous study (ADS10) to 
investigate the generation process of magnetic fields 
due to the ionizing radiation from first stars with more precision. 
We take into consideration 
not only the effects of radiation force but also the
Biermann battery mechanism, 
utilizing the two-dimensional radiation hydrodynamics simulations.
Then we discuss whether the magnetic field strength obtained in our
study could be important for subsequent star formation process.
In section 2, we describe the basic equations and the setup of our model.
We show the results of our calculations in section 3.
Sections 4 and 5 are devoted to the discussions and summary.
\section{Basic equations \& Model}
\subsection{Equation of magnetic field generation}
\label{mag_generation}
According to ADS10, the equation of magnetic field generation is given as
\begin{eqnarray}
\dpdp{\bm{B}}{t}=\bm{\nabla}\times\left(\bm{v}\times\bm{B}\right)
-\frac{c}{e n_{\rm e}^2}\bm{\nabla}n_{\rm e}\times\bm{\nabla}p_{\rm e}
-\frac{c}{e}\bm{\nabla}\times\bm{f}_{\rm rad}\label{eq:B_growth}
\end{eqnarray}
where $\bm{B}$, $\bm{v}$ and $e$ denote the magnetic flux density, fluid velocity,
and the elementary charge, respectively.
$p_{\rm e}$ and $n_{\rm e}$ represent the pressure and the number density of electrons.
$\bm{f}_{\rm rad}$ is the radiation force acting on an electron.
The first term on the right-hand side is the advection term of magnetic
flux, whereas the second term describes the Biermann battery term
\citep{biermann}, which was not included in our previous study (ADS10).
The third term is the radiation term which represents the momentum
transfer from photons to gas particles. Remark that the
dissipation term due to the resistivity of the gas is omitted, since it
is negligible in comparison with the other terms (ADS10).

\subsection{Radiation force and photoionization rate}
The radiation force on an electron, $\bm{f}_{\rm rad}$ involves two
processes.
The first one is the contribution by the Thomson scattering, $\bm{f}_{\rm rad,T}$.
$\bm{f}_{\rm rad,T}$ is given by
\begin{eqnarray}
\bm{f}_{\rm rad,T}  = \frac{\sigma_{\rm T}}{c}\int_0^{\nu_{L}}\bm{F}_{0\nu}d\nu 
+ \frac{\sigma_{\rm T}}{c}\int_{\nu_{L}}^{\infty}\bm{F}_{0\nu}\exp\left[- \tau_{\nu_{L}}a\left(\nu\right)\right]  d\nu, \nonumber\\
\label{eq:momtr_T}
\end{eqnarray}
where $\sigma_{\rm T}$ denotes the cross section of the Thomson scattering,
$\bm{F}_{0\nu}$ is the unabsorbed energy flux density,
$\nu_{L}$ is the Lyman-limit frequency,
$\tau_{\nu_{L}}$ denotes the optical depth at the Lyman limit regarding the photoionization,
and $a\left(\nu\right)$ is the frequency dependence of photoionization
cross section which is normalized at the Lyman limit frequency.

Another source of the radiation force is the momentum transfer
from photons to electrons through the photoionization process.
The force expressed as $\bm{f}_{\rm rad,I}$, is
\begin{eqnarray}
\bm{f}_{\rm rad,I}  = \frac{1}{2}\frac{n_{\rm HI}}{c n_{\rm e}}
\int_{\nu_{L}}^{\infty} \sigma_{\nu_{L} } a\left(\nu\right)
\bm{F}_{0\nu}\exp\left[- \tau_{\nu_{L}} a\left(\nu\right)\right]  d\nu,\label{eq:momtr_ion}
\end{eqnarray}
where $\sigma_{\nu_{L}}$ is the photoionization cross section at the
Lyman limit and $n_{\rm HI}$ represents the number density of neutral
hydrogen atoms.

In order to assess the electron number density, we also solve the following photoionization rate equation for electrons:
\begin{eqnarray}
\frac{\partial y_{\rm e}}{\partial t} + (\bm{v} \cdot \bm{\nabla}) y_{\rm e}
 &=& \frac{\Gamma}{n_{\rm H}} - \alpha_{\rm B} y_{\rm e} y_{\rm p} n_{\rm H} + k_{\rm coll}y_{\rm e}y_{\rm HI} n_{\rm H},\label{eq:cont_e}
 \label{eq:cont_p}
\end{eqnarray}
where $\alpha_{\rm B}$ and  $k_{\rm
coll}$ denote the case B recombination rate and collisional ionization
rate per unit volume, respectively.  
 $y_{\rm e}$, $y_{\rm p}$ and $y_{\rm HI}$ is the number fraction of
 electrons, protons and neutral hydrogen atoms, respectively.
$n_{\rm H}$ is the number density of hydrogen nuclei and $\bm{v}$
denotes the velocity of the fluid.
The photoionization rate per unit volume, $\Gamma$, is also obtained by the formal solution of the radiation transfer equation:
\begin{eqnarray}
\Gamma = n_{\rm HI}\int_{\nu_{L}}^{\infty}\sigma_{\nu_{L}}
\frac{{F}_{0\nu}}{h\nu}\exp\left[- \tau_{\nu_{L}}a\left(\nu\right)\right]  d\nu.
\end{eqnarray}
%In this paper, the advection term (second term) on the left-hand side of equation
%(\ref{eq:cont_e}) is omitted, since this term is $\la \sim 10^{-2}$ time
%smaller than the other terms throughout the simulation.

\subsection{Hydrodynamics with heating/cooling}
We solve the ordinary set of hydrodynamics equations:
\begin{eqnarray}
  \frac{D\rho}{Dt}&=&-\rho \bm{\nabla}\cdot \bm{v},\label{eq:continuity_eq}\\
  \frac{D\bm{v}}{Dt}&=&-\frac{1}{\rho}\bm{\nabla}p-\bm{\nabla}\phi_{\rm DM},\label{eq:motion_eq}\\
  \rho\frac{D\epsilon}{Dt}&=&G-L-p\bm{\nabla}\cdot\bm{v},\label{eq:energy_eq}
\end{eqnarray}
and the equation of state
\begin{eqnarray}
  p=(\gamma-1)\rho \epsilon,\label{eq:state_eq}
\end{eqnarray}
where $\rho$, $p$, and $\epsilon$ are 
the density, pressure, and the specific energy of the fluid, respectively.
$\gamma$ denotes the specific heat ratio.
$G$ and $L$ are the radiative heating rate and cooling rate per unit volume, respectively.
$\phi_{\rm DM}$ is the gravitational potential of the dark matter halo(see \ref{sec:setup}).
The feedback from magnetic fields to the fluid is neglected in equation(\ref{eq:motion_eq}), since we consider the generation of very weak magnetic fields.
We also omit the self gravitational force of gas, which is unimportant as long as we consider the gas with $n_{\rm H} \la 10^3{\rm cm}^{-3}$.

In order to perform hydrodynamics simulations, we use the Cubic Interpolated
Profile (CIP) method \citep{yabe}.
 The CIP scheme basically tries to solve not
only the advection of physical quantities but also the derivatives of
the quantities. Using this scheme, we can capture spatially sharp profiles
of fluids, that are always expected in the problems including the
propagation of ionization fronts. 
In this paper, the CIP scheme is applied to the advection terms of
equations (\ref{eq:cont_p}) and (\ref{eq:continuity_eq})-(\ref{eq:energy_eq}).

Using the formal solution of radiation transfer equations, the radiative
heating rate $G$ is assessed as 
\begin{eqnarray}
  G=n_{\rm HI}\int_{\nu_{L}}^{\infty}\frac{F_{0\nu}}{h\nu}h\left(\nu-\nu_{L}\right)
  \sigma_{\nu_{L}}a\left(\nu\right)\exp\left[- \tau_{\nu_{L}}a\left(\nu\right)\right]d\nu.\nonumber\\
\label{eq:heating_rate}
\end{eqnarray}
The cooling rate, $L$, is given as
\begin{eqnarray}
  L=L_{\rm coll}n_{\rm e}n_{\rm HI}+L_{\rm rec}n_{\rm e}n_{\rm p}
  +L_{\rm exc}n_{\rm e}n_{\rm HI}+L_{\rm ff}n_{\rm e}n_{\rm p},\nonumber\\\label{eq:cooling}
\end{eqnarray}
where $L_{\rm coll}$, $L_{\rm rec}$, $L_{\rm exc}$ and $L_{\rm ff}$ are
 the cooling coefficients regarding the collisional-ionization,
 the recombination, 
 the collisional excitation and
 the free-free emission.
These cooling rates are taken from the compilations in \citet{fukugita_kawasaki}.

\begin{figure}
  \begin{center}
    \includegraphics[width=12cm]{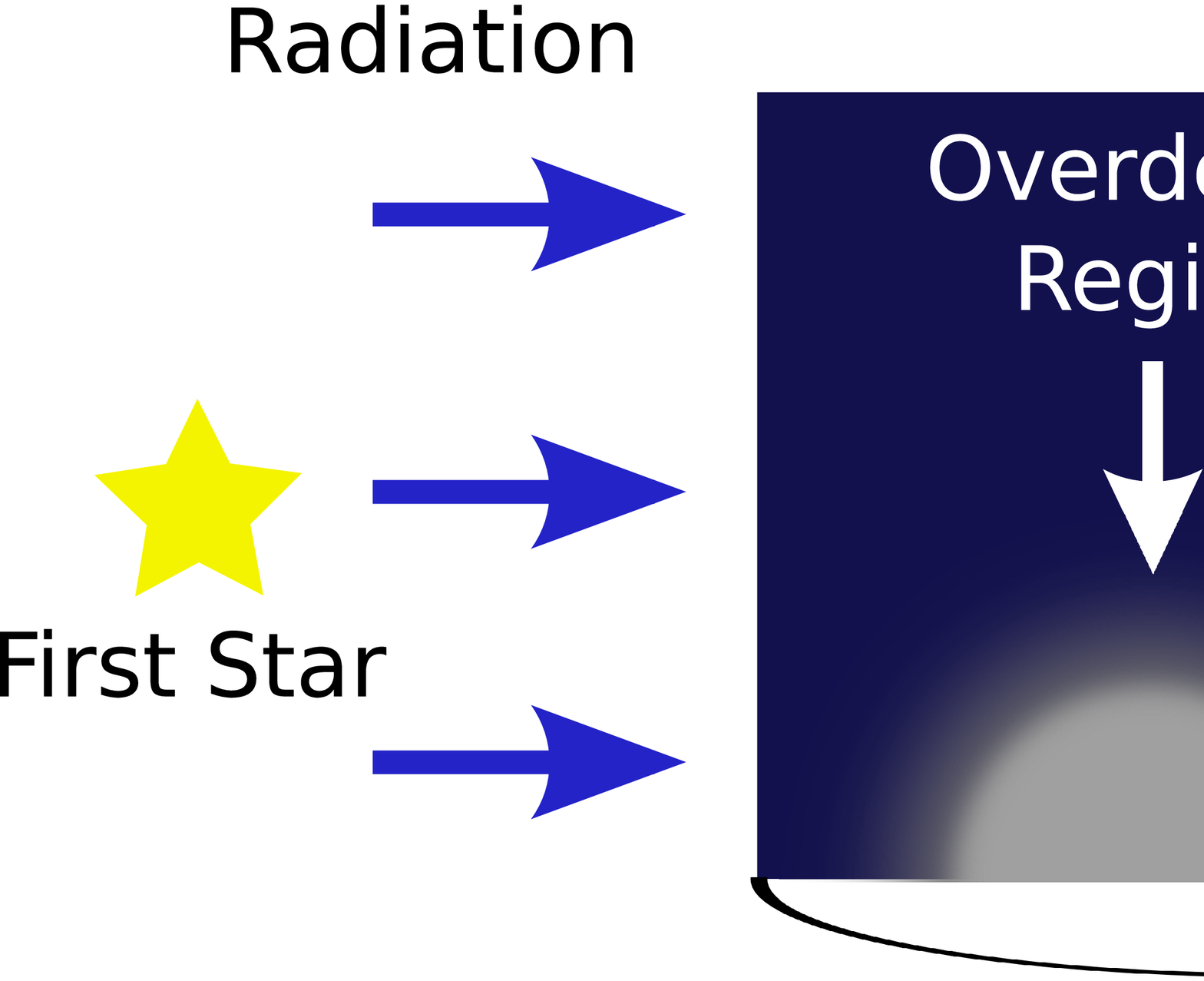}
  \end{center}
  \caption{Schematic view of the computational domain. The shadowed
    region is formed behind the over dense region , while the other region
    is exposed to radiation field.}\label{fig1}
\end{figure}
\subsection{Setup}\label{sec:setup}
We consider a mini-halo in the IGM at redshift $z \simeq 20$ exposed to an intense radiation flux from a nearby first star.
We assume that the dark matter density profile of the mini-halo is
described by  the NFW profile \citep{navarro97}
\begin{eqnarray}
\rho_{\rm DM}\left( r \right) = \frac{\rho_{\rm s}}{\left(r/r_{\rm s}\right) \left(1+r/r_{\rm s}\right)^2} ,\label{eq:nfw}
\end{eqnarray}
where $r$ is the radial destance from the center of halo. $\rho _{\rm s}$ and $r_{\rm s}$ are a characteristic density and radius, which are determined by a halo mass $M_{\rm halo}$ collapsing at redshift $z$ \citep{prada11}.
We assume the gas density profile is a core-halo structure with  $\rho \propto r^{-2}$ envelope.
The core radius $r_{\rm c}$ is determined by the core density and the total gas mass $\left(M_{\rm gas}=\left(\Omega_{\rm B}/\Omega_{\rm M}\right)M_{\rm halo}\right)$.
We study various cases of core densities $\left(n_{\rm 0}{\rm cm}^{-3}\right)$, the halo mass $M_{\rm halo}$ and distances between the source star and halo center $\left(D{\rm kpc}\right)$.
We assume  stationarity of the mini-halo with respect to the source star.
This assumption is based upon the fact that the the
change of distance between the star and the halo due to the relative
motion is smaller than $D$, if we consider the cosmic
expansion. 
We remark that in case the neighboring overdense region and the
source star is contained in a same halo of $>10^7M_\odot$, change of the
distance due to the velocity dispertion of the halo could have significant effect especially for low mass source stars.

Initially, the ambient gas is assumed to be neutral when the source star is turned on.
The initial number density of the ambient gas in the intergalactic space is
$n_{\rm IGM} = 10^{-2}{\rm cm}^{-3}$.
We assume that the initial temperature is $500{\rm K}$.
We perform two-dimensional simulations in the cylindrical coordinates
assuming axial symmetry.
As shown in Figure \ref{fig1}, the computational domain is 100 pc
$\times$ 200 pc in $R-z$ plane. We consider source stars of various masses, 
$M_{\ast}=500$,$300$,$120$,$60$, $25$ and $9 M_{\odot}$. The luminosities,
effective temperatures and the ages of these stars are taken from the table
of \citet{schaerer}.
The incident radiation from the source star is assumed to be
perpendicular to the left edge of the computational domain.

We employ four models (A-D) of different $M_{\rm halo}, $$n_0$ and $D$ listed in Table
\ref{tab1}, these are plausible values of the minihalos in standard
$\Lambda$CDM cosmology(see section \ref{DepnD}). 
The number of grids we use in these simulations is basically
$250\times 500$ which is confirmed to be enough to obtain physical
results by the convergence study (see section \ref{convergence}).

\section{Results}
\subsection{Typical results}
\begin{figure}
  \begin{center}
    \includegraphics[width=11cm]{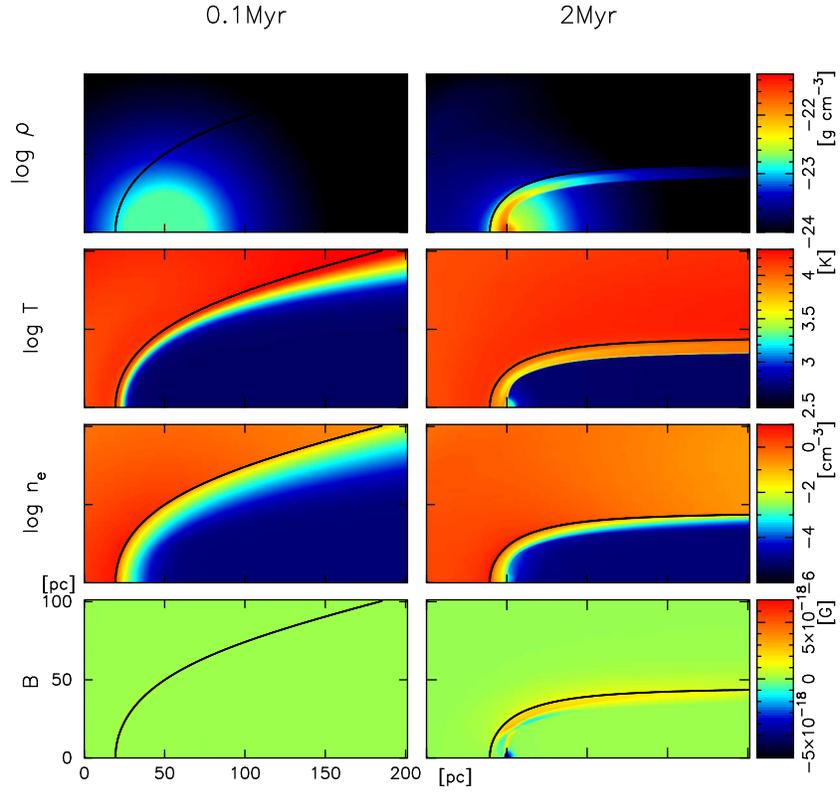}
  \end{center}
  \caption{Two snapshots (left:0.1Myr, right:2Myr) for the model A are
 shown. Four rows correspond to the mass density of gas [${\rm g
 cm^{-3}}$](top),   
  the gas temperature $[\rm K]$, the number density of electron [${\rm 
 cm^{-3}}$],  and the magnetic field strength [$G$], respectively. 
 Black solid lines represent the position of
 the ionization fronts.}
  \label{fig2}
\end{figure}

First, we consider a first star of 500 $M_{\odot}$($t_{\rm age}=2\times 10^{6}{\rm yr}$).
Figure \ref{fig2} shows the results for the model A. Two columns
correspond to the snapshots at $0.1$Myr and $2$Myr.
The top row shows the color contour of the mass density of the gas. The gas
density distribution is isotropic at $0.1$Myr, while it is highly
disturbed by the radiation flux from the left at $2$Myr. We also can find a
shock front at $2$Myr inside the core of the overdense region.

The second and third rows show the maps of the temperature and the electron
density. Clearly, an ionization front is generated by the flux from the
left, and it propagates into the core of the cloud. It is also worth
noting that the pattern of temperature and electron density distribution
are similar, but not identical with each other. 
Slight difference between the two
contour maps directly leads to the nontrivial Biermann battery term. 

The bottom row shows the distribution of the generated magnetic field
strength. As expected, we obtain strong magnetic fields in the neighborhood of the
ionization front. 
In addition, strong magnetic fields is produced at the center of core where the density is the highest.
The peak field strength is as strong as $6\times 10^{-18}$G in this case.

\begin{figure}
  \begin{center}
    \includegraphics[width=12cm]{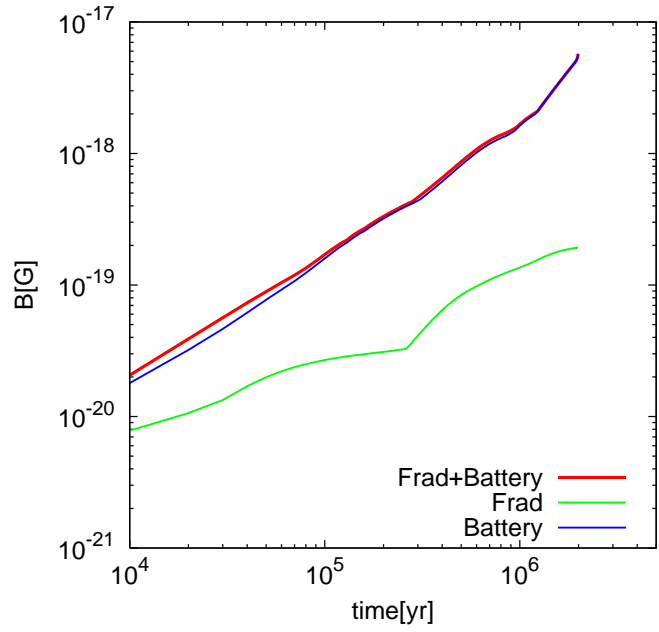}
  \end{center}
  \caption{Time evolution of the peak magnetic field for the model A
 (red). Other two curves correspond to the field generated by Biermann
 battery (blue) and radiation pressure (green). }
  \label{fig3}
\end{figure}

In Figure \ref{fig3}, we show the time evolution of the peak magnetic
field strength in the simulated region (red curve). We also plot the
peak magnetic field generated by the Biermann battery term (blue) as well
as the one by the radiation processes (green).
The field strength grows almost linearly in this model, and the final
strength is as large as $\sim 6\times 10^{-18}$G. 
We also find that the radiation process is less important than the
Biermann battery effect.
However, it is still noteworthy that the difference between the two
contributions is only a factor of $\sim 10$, although the nature of these
processes are very different from each other.
  
\subsection{Dependence on $n_0$, $D$, $M_{\ast}$}
\label{DepnD}
\begin{figure}
  \begin{center}
    \includegraphics[width=12cm]{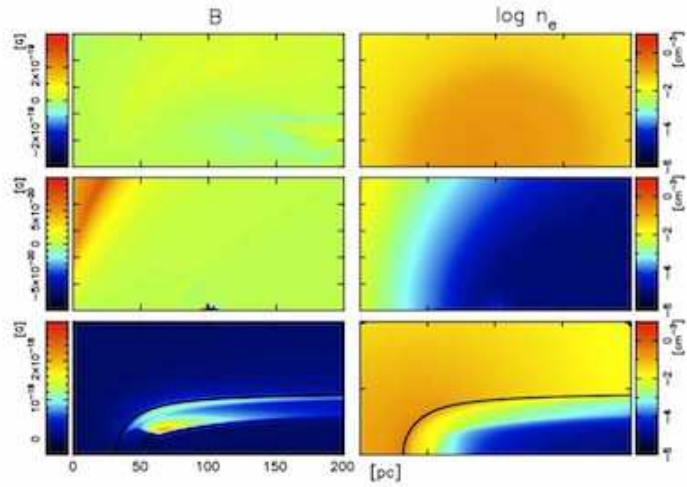}
  \end{center}
  \caption{Snapshots at 2Myr for the models B,C and D are shown. Left column : magnetic
 filed strength, $B$[G] . Right column: electron density, $n_{\rm e}$[cm$^{-3}$].}
  \label{fig4}
\end{figure}

We also investigate the other models of $n_0$ and $D$ listed in Table \ref{tab1}.
The snapshots at 2Myr of these models (B,C and D from top to bottom) are shown in Figure \ref{fig4}. 
The left column shows the magnetic field strength, whereas the right
column illustrates the number density of electrons.
In the model B, the generated magnetic field is of the order of $\sim
10^{-19}$G (top left panel),
which is smaller than that in the model A.
Since the ionization front is not trapped by the dense core in the model B
(top right), the source terms of equation (\ref{eq:B_growth})
become very small as soon as the ionization front passes through the
core. As a result, the magnetic field does not have enough time to grow.
In the models C and D,  the core is 10 times more distant from the
source star than the models A and B.
Therefore, the ionization front is trapped at lower density regions, and
becomes less sharp than that in model A.
Consequently, the generated magnetic field is smaller than that in the
model A (middle and bottom row).

\begin{figure}
  \begin{center}
    \includegraphics[width=12cm]{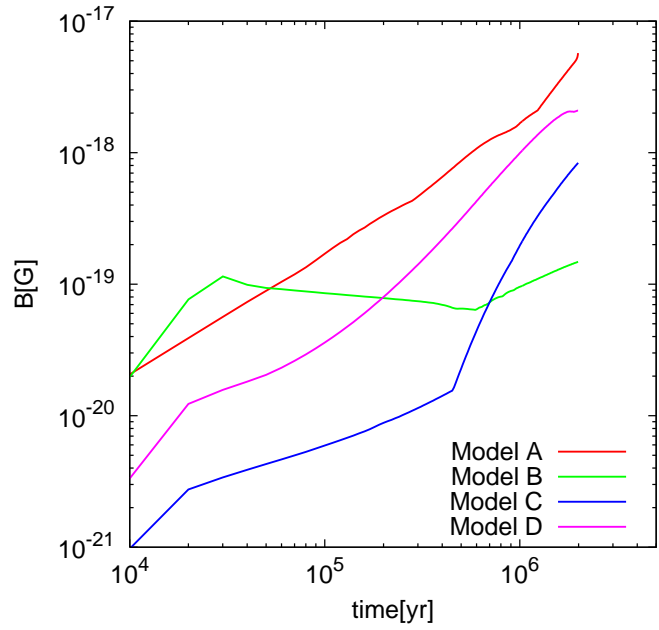}
  \end{center}
  \caption{Time evolution of the peak magnetic field for the models A-D.}
  \label{fig5}
\end{figure}

We also plot the time evolution of the peak field strengths of the models A-D
in Figure \ref{fig5}. The models A, C and D mostly increase monotonically, whereas the model B has a clear plateau/decline. Such different behavior
also comes from the fact that the ionization front immediately passes
through the core in the model B. In such case, the time for the magnetic
field to grow is not enough as stated above. In addition, the fluids that host
the generated magnetic fields expand due to the thermal
pressure of photoionized gas. Such expansion results in the slight decline of
the magnetic field strength.

\begin{figure}
  \begin{center}
    \includegraphics[width=12cm]{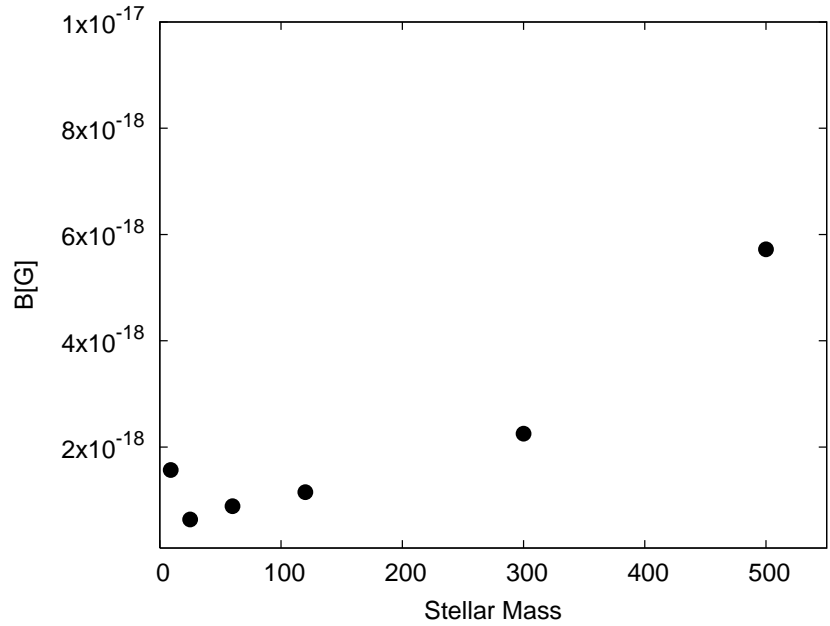}
  \end{center}
  \caption{The maximal magnetic field strength for $D=200$pc,
 $n_0=10{\rm cm^{-3}}$, with various $M_{\ast}$.}
  \label{fig6}
\end{figure}

The dependence on the source stellar mass $M_\ast$ is also studied.
We employ six models of $M_{\ast}=500$,$300$,$120$,$60$,$25$ and
$9M_{\odot}$, while the other parameters are same as the model A.
Figure \ref{fig6} shows the peak magnetic field strength as a function
of $M_{\ast}$. The peak field strength of each run is evaluated
when it gets to the time for the death of the source star.
The peak field strength basically increases as the stellar mass becomes
more massive. However, the difference between the field strength of
$M_{\ast}=9M_\odot $ and $500M_\odot$ is a factor of 9, which is
a small difference for such a mass difference.
The reason of this behavior comes from the fact that
the more massive the stars are, the more ionizing photons they emit, but
also the shorter lifetime they have.  These two competing effects cancel
with each other. 

In any case, the generated magnetic field strengths stay around
$10^{-18}-10^{-17}$G, if we consider the stellar mass of $9 M_\odot$
- $500 M_\odot$, which range is wide enough for first stars. 

The parameters $n_0$ and $D$ we employ in this paper are reasonable values.
The baryonic density of the virialized halo at $z=20$ is $\sim1{\rm cm}^{-3}$ ,
and the sizes of the first halos are $\la 100$pc. In addition, the radii
of cosmological HII regions in the numerical simulations are a few kpc
\citep[e.g.,][]{yoshida07}. Thus, our models A-D are reasonable for the 
standard cosmological model.

To summarize the results of this section (\ref{DepnD}), 
the magnetic field strength generated around a first star is  
$10^{-19}{\rm G} < B <  10^{-17}$G for $M_{\ast}=500 M_\odot$, and a
factor of a few smaller for less massive stars. 

\subsection{Coherence length}
In addition, we also
remark that the coherence length of the magnetic field.
Due to the limited computational resource, we employ rather small box
size ($\sim 200$pc). The resultant coherence length of the magnetic
field clearly exceeds the box size in most of the cases. In addition, if
the ionization structure is more or less similar to that of our previous
results in ADS10, the coherence length will be as large as $\sim 1$kpc,
which is the size of the shadow.

\subsection{Convergence check}
\label{convergence}
\begin{figure}
  \begin{center}
    \includegraphics[width=12cm]{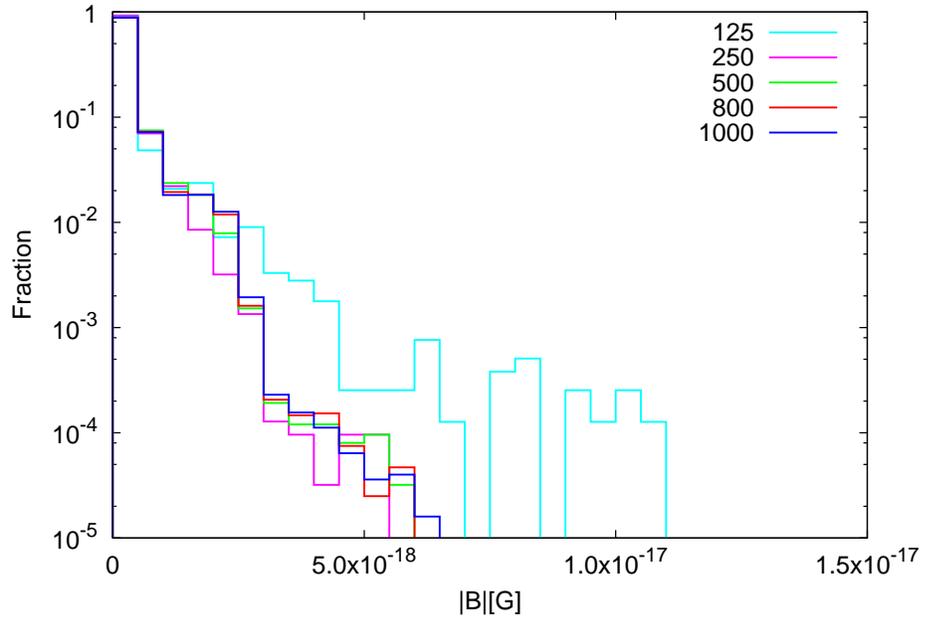}
  \end{center}
  \caption{Magnetic field probability distribution functions are plotted
 for various $N_Z$.}
  \label{fig7}
\end{figure}

Since the equation of the magnetic field generation (\ref{eq:B_growth})
 includes spatial derivatives in its source terms, we have to pay
 attention to the effects of the cell size on our numerical results.
 We check the numerical convergence of the magnetic field
 strengths for the model A. We perform runs with $N_R\times N_z=63\times125,125\times250,250\times500({\rm
 canonical}),400\times800,500\times1000$, where $N_R$,$N_Z$ is the number of grids in $R$-axis and $z$-axis, respectively.
In Figure \ref{fig7}, the magnetic field probability
 distribution functions (PDFs) of these runs are plotted. The axis of abscissas
 is the absolute value of the magnetic field strength, while the vertical axis represents the
 fraction of grid cells that fall in the range of $[B,B+\Delta B]$,
 where $\Delta B = 5 \times 10^{-19}$G. 
  The peak field strength decline as the number of grids increases.
  However, the peak field strength converges $\sim 6 \times 10^{-18}$G 
  and PDFs show similar distributions for $N_R\times N_Z\ge 250\times500$.
  Therefore, the numerical results of the simulations converge very well
  at $N_R\times N_Z=250\times500$, 
  which we use in other runs. 

\section{Discussions}
In this paper, we investigate the magnetic field generated by first stars.
As a result, the maximal magnetic field strength is $\la 10^{-17}$G,
mainly generated by the Biermann battery mechanism.

In fact, the order of magnitude of the magnetic field generated by the
Biermann battery could be assessed as
\begin{eqnarray}
&&B \sim  \frac{c}{n_{e}^{2} e} \left(\frac{n_{e}}{\Delta r} \right)
 \left(\frac{p_{e}}{\Delta r} \right)\sin\theta \; t_{\rm age} \nonumber\\ 
 &&\sim 5.0 \times 10^{-17}{\rm G}\left( \frac{t_{\rm age}}{2{\rm Myr}} \right) \left(\frac{\sin\theta}{0.1}\right)
 \left(\frac{\Delta r}{1 {\rm pc}} \right)^{\hspace{-2mm}-2}\hspace{-2mm}\left(\frac{T}{10^{4}{\rm K}}\right),\label{eq:bb_order}
\end{eqnarray}
where $\Delta r$ denotes the typical length of $n_{\rm e}$ and $p_{\rm e}$ change significantly,
$\theta$ is the angle between $\bm{\nabla}n_{\rm e}$ and $\bm{\nabla}T$.
The resultant value of above equation is roughly consistent with the
results of our numerical simulations. 

We find the relative importance of the Biermann battery effect versus the
radiative processes in this paper. However, this is only relevant for
present setup, because the two effects depend differently on various
parameters. In particular, the Biermann battery effect do not depend on
the flux of the source star directly (see eq.\ref{eq:bb_order}), 
whereas the radiation force
is proportional to the flux. This means if consider the magnetic
field generation process in the very neighbor of the source objects,
such as the accretion disks of the protostars/black holes, radiative
processes could play central roles in the generation of magnetic
field. We will study this issue in the near future.

The field strength obtained in this paper is similar to the results of
\citet{xu}, in which they investigated magnetic fields in collapsing
mini-halos/prestellar cores.
If we assume that the magnetic field generated around first stars are
brought into another prestellar core, and evolve during the collapse of
the primordial gas in a similar way to \citet{xu},
the magnetic field will be amplified up to 
$\sim10^{-13}$ G at $10^3{\rm cm}^{-3}$.
However, this magnetic field will not affect subsequent star formation,
since the magnetic field strength required for jet
formation\citep{machida06} and MRI activation\citep{tan_blackman} is
$10^{-10}-10^{-9}$G at $10^{3}{\rm cm}^{-3}$, if we assume simple flux
freezing condition.
On the other hand, recent studies suggest that weak seed magnetic field is
amplified by the turbulence during the first star formation\citep{schleicher10,sur10}.
In \citet{sur10}, weak seed magnetic fields are exponentially amplified
by small-scale dynamo action if they employ sufficient numerical resolutions. 
In this case, the magnetic fields generated by the first stars in this paper might be amplified and affect subsequent star formation.
However, these studies assume a priori given turbulence and initial
magnetic field. To try to settle this issue, we need cosmological MHD simulations with
very high resolution.

\section{Summary}
In summary, we have investigated the magnetic field generation process
by the radiative feedback of first stars, including the effects
of radiation force and the Biermann battery. As a result, we found
$10^{-19}{\rm G }\la B\la 10^{-17}$G on the boundary of the shadowed
region, if we take reasonable parameters expected from standard theory
of cosmological structure formation. The resultant field strength with a
simple assumption of flux freezing suggests that the such magnetic field
is unimportant for the star formation process. However, it could be
important if the magnetic field is amplified by turbulent motions of the
star forming gas could.

\bigskip
We appreciate the anonymous referee for helpful comments.
We also thank N.Tominaga and M.Ando for fruitful discussions. 
This work was supported by Ministry of Education, Science, Sports and
Culture, Grant-in-Aid for Scientific Research (C), 22540295.

\begin{table}
\begin{center}
\begin{tabular}[t]{|c|c|c|c|c|}
\hline
Model&A&B&C&D\\ \hline
$M_{\rm halo}[h^{-1}M_{\odot}]$&$5\times10^6$&$1\times10^5$&$5\times10^6$&$1\times10^5$\\
$n_0 [{\rm cm^{-3}}]$&10&1&10&1\\
$D [{\rm kpc}]$&0.2&0.2&2&2\\
Magnetic field [G]&$5.7\times10^{-18}$&$1.5\times10^{-19}$&$8.4\times10^{-19}$&$2.1\times10^{-18}$\\
\hline
\end{tabular}
\end{center}
\caption{The peak magnetic field strengths of four models of $M_{\ast}=500M_\odot$.}
\label{tab1}
\end{table}

\end{document}